\pdfoutput=1
\documentclass[11pt]{article}
\usepackage[margin=2.6cm]{geometry}
\usepackage[english]{babel} 
\usepackage[utf8]{inputenc}
\usepackage{indentfirst}
\usepackage{amsmath}
\usepackage{amssymb}
\usepackage{url}
\usepackage{graphicx}
\usepackage{braket}
\usepackage{slashed}
\usepackage{cancel}
\usepackage[font=footnotesize,labelfont=bf]{caption}
\usepackage[usenames, dvipsnames]{xcolor}
\usepackage{booktabs}
\usepackage{dsfont}
\usepackage{pdflscape}
\usepackage{float}

\usepackage[colorlinks=true, urlcolor=MidnightBlue, linkcolor=MidnightBlue, citecolor=Mahogany, pdfborder={0 0 0}]{hyperref}
\usepackage[sort&compress,numbers]{natbib}
\usepackage{doi}
\usepackage[capitalise]{cleveref}

\newcommand{\CP}{$C\!P$}
\newcommand{\CPH}{$C\!P_{\textrm{H}}$}
\newcommand{\CPL}{$C\!P_{\textrm{HL}}$}

\newcommand{\SLV}[1]{\boldsymbol{#1}}

\DeclareTextFontCommand{\comm}{\color{MidnightBlue}\em}

\title{\Large\bfseries{\CP{} violation in $\eta$ muonic decays}}
\author{\normalsize{
        Pablo Sanchez-Puertas{\color{Mahogany}\thanks{psanchez@ifae.es}}$^{\color{Mahogany}{\  a,b}}$
        } \vspace{0.2cm}  \\ 
        {\small{$^{\color{Mahogany}{a}}$\textit{Institut de F{\'i}sica d'Altes Energies (IFAE),}}} \\
        {\small{\textit{The Barcelona Institute of Science and Technology,}}} \\
        {\small{\textit{Universitat Aut{\'o}noma de Barcelona, E-08193 Bellaterra (Barcelona), Spain,}}}\\
        {\small{\textit{$^{\color{Mahogany}{b}}$Faculty of Mathematics and Physics, Institute of Particle and Nuclear Physics,}}} \\
        {\small{\textit{Charles University in Prague, V Holešovičkách 2, 18000 Praha 8, Czech Republic.}}}
}
\date{}

\begin{document}\renewcommand{\abstractname}{\vspace{-\baselineskip}} \maketitle

\begin{abstract} 

In this study, we investigate the imprints of \CP{} violation in certain $\eta$ muonic decays that could arise within the Standard Model effective field theory. In particular, we study the sensitivities that could be reached at REDTOP, a proposed $\eta$ facility. After estimating the bounds that the neutron EDM places, we find still viable to discover signals of \CP{} violation measuring the polarization of muons  in $\eta\to\mu^+\mu^-$ decays, with a single effective operator as its plausible source. 

\end{abstract}

\section{Introduction}

In this article, we investigate the signal of \CP{} violation in a set of $\eta$ muonic decays. In doing so, we assume the signal as arising from heavy physics, so that the Standard Model effective field theory (SMEFT) can be applied. As a result, two different scenarios arise: that of \CP{}-violating purely hadronic operators (\CPH{}) and that of \CP{}-violating quark-lepton ones (\CPL{}). We provide the required amplitudes for Monte Carlo (MC) generators and evaluate the impact of such operators in terms of certain asymmetries, using as a benchmark a proposed $\eta$ factory with the ability of measuring the polarization of muons: REDTOP~\cite{Gatto:2016rae}. After estimating the impact of these operators on the neutron dipole moment (nEDM), we find that \CP{}-violating quark-lepton interactions could be at the reach of REDTOP, while evading nEDM bounds. Dealing with muons, these bounds are complementary to those of the electron case which have been put recently after the ACME Collaboration results~\cite{Andreev:2018ayy} in Ref.~\cite{Cesarotti:2018huy}.

The article is organized as follows: in \cref{sec:cpscen}, we discuss the two \CP{}-violating scenarios arising from the SMEFT operators and their connection from quark to hadron degrees of freedom (e.g.~with $\eta$-physics). Then, in \cref{sec:mudec}, we compute the $\eta\to\mu^+\mu^-, \gamma\mu^+\mu^-, \mu^+\mu^-e^+e^- $ decays, accounting for the polarization of muons in dilepton and single-Dalitz decays. We provide the required expressions for MC generators and compute different asymmetries that could be generated, providing the sensitivities at reach at REDTOP. Finally, in \cref{sec:nedm}, we evaluate the impact of both \CP{}-violating scenarios on the nEDM, that sets stringent constraints.

\section{The \CP{}-violating scenarios}\label{sec:cpscen}

In our study, we assume that the \CP{}-violating new-physics effects are heavy enough to be described through the SMEFT. Following the operator basis in Ref.~\cite{Grzadkowski:2010es}, we can find here \CP{} violation in three different sectors: that involving the hadronic part only, which we include in the \CPH{} category; that mixing quark and leptons, which we include in the \CPL{} one; and that affecting lepton-photon interactions ($\mathcal{O}_{eW,eB}$), which we checked to be negligible and discard\footnote{Particularly, the $\mu$EDM~\cite{Tanabashi:2018oca} put tight constraints on these operators~\cite{Pruna:2017tif}.} for brevity.\footnote{We could have \CP{} violation in the pure leptonic side via the $\mathcal{O}_{le}$ operator~\cite{Pruna:2017tif,Panico:2018hal}. However, this is irrelevant in what we find the best channel, $\eta\to\mu^+\mu^-$, and should be negligible in other cases---see discussions later.
}
Regarding the \CPH{} category, since we are interested in decays connecting $\eta$ to muons, these will result, at low-energies, in a \CP{}-violating shift of the $\eta\gamma^*\gamma^*$ coupling, 
a scenario extensively discussed in the literature in the context of light pseudoscalar mesons~\cite{Samios:1962zza,Uy:1990hu,Abouzaid:2008cd}. In general, one has
   \begin{multline}\label{eq:ggvertex}
      i\mathcal{M}^{\mu\nu} = ie^2\Big(\epsilon^{\mu\nu\rho\sigma}q_{1\rho}q_{2\sigma}F_{\eta\gamma^*\gamma^*}(q_1^2,q_2^2) 
                     + \left[g^{\mu\nu}(q_1\cdot q_2) -q_2^{\mu}q_1^{\nu}\right] F_{\eta\gamma^*\gamma^*}^{C\!P1}(q_1^2,q_2^2) \\
                     + \left[ g^{\mu\nu}q_1^2q_2^2 -q_1^2 q_2^{\mu}q_2^{\nu} -q_2^2 q_1^{\mu}q_1^{\nu} 
                                    +(q_1\cdot q_2)q_1^{\mu}q_2^{\nu} \right] F_{\eta\gamma^*\gamma^*}^{C\!P2}(q_1^2,q_2^2) \Big), 
   \end{multline}
where $\epsilon^{0123}=+1$, $F_{\eta\gamma^*\gamma^*}$ is the standard transition form factor (TFF), and the latter two are \CP{}-violating ones. For some details on our TFF description, we refer to \cref{sec:TFF}. Accounting for the hadronization details linking the SMEFT \CPH{} operators to $F_{\eta\gamma^*\gamma^*}^{C\!P1,2}(q_1^2,q_2^2)$ in a quantitative manner is a formidable task; however, this is enough to our purposes as we shall see.
Coming back to the \CPL{} category, the relevant operators here are
  \begin{align}
    \mathcal{O}_{\ell equ}^{(1)} =&{} \frac{c_{\ell equ}^{(1)prst}}{v^2}(\bar{\ell}_p^j e_r)\epsilon_{jk}(\bar{q}^k_s u_t)  +\operatorname{h.c.}
         \!\!\!\!&\!\! &\to-\frac{\operatorname{Im}c_{\ell equ}^{(1)prst}}{2v^2}
               \left[ (\bar{e}_pi\gamma^5e_r)(\bar{u}_su_t) +(\bar{e}_pe_r)(\bar{u}_si\gamma^5u_t) \right], \\
    \mathcal{O}_{\ell edq} =&{} \frac{c_{\ell edq}^{prst}}{v^2}(\bar{\ell}^j_p e_r)(\bar{d}_s q^j_t) +\operatorname{h.c.}
        \!\!\!\!&\!\! &\to \frac{\operatorname{Im}c_{\ell edq}^{prst}}{2v^2}
               \left[ (\bar{e}_pi\gamma^5e_r)(\bar{d}_sd_t) -(\bar{e}_pe_r)(\bar{d}_si\gamma^5d_t)\right],
  \end{align}
where $v^2\simeq \sqrt{2} G_F$, and $\{p,r,s,t\}$ flavor indices. Concerning the $\eta$, these produce \CP{}-violating interactions of the kind $\mathcal{L} = -\mathcal{C}(\eta \bar{e}^p e^r)$, with\footnote{The FKS scheme~\cite{Feldmann:1999uf} is employed for describing the $\eta-\eta'$ mixing (with parameters from Ref.~\cite{Escribano:2015yup}), implying $\bra{\Omega} 2m_a \bar{q}^ai\gamma^5q^a \ket{\eta(P)} = F_{\eta}^q m_{\pi}^2 \operatorname{tr}(a\lambda^q) + F_{\eta}^s (2m_K^2 - m_{\pi}^2) \operatorname{tr}(a\lambda^s)$. Moreover, we employ the quark mases in PDG~\cite{Tanabashi:2018oca}, at the scale $\mu=2$~GeV: $m_u\simeq m_d\equiv\hat{m}=3.5$~MeV and $m_s=96$~MeV.} 
   \begin{align}
      \mathcal{C} \equiv \frac{\operatorname{Im}c_{\mathcal{O}}}{2v^2} \bra{\Omega} \bar{q}^si\gamma^5 q^t \ket{\eta} = 
       \operatorname{Im}(1.57(c_{\ell equ}^{(1)pr11}+c_{\ell edq}^{pr11}) -2.37  c_{\ell edq}^{pr22})\times 10^{-6}. \label{eq:mathcalC}
   \end{align}

\section{Muonic decays and asymmetries}\label{sec:mudec}

Having discussed the relevant hadronic matrix elements, we are prepared to discuss the different muonic decays, which in the first two cases involve the muon polarization---a property that can be measured at REDTOP~\cite{Gatto:2016rae}.

\subsection{The golden channel: $\eta\to\mu^+\mu^-$}\label{sec:dilepton}

In general, there are two structures governing the $\eta\to\mu^+\mu^-$ decay amplitude, namely~\cite{Martin:1970ai,Ecker:1991ru} ($g_P$ and $g_S$ are dimensionless):
  \begin{equation}\label{eq:gencoup}
    i\mathcal{M} = i \left[ g_P (\bar{u}i\gamma^5 v) + g_S (\bar{u}v) \right], 
  \end{equation} 
where the first(second) term is \CP{} even(odd). In the SM, $g_P=-2m_{\mu}\alpha^2 F_{\eta\gamma\gamma}\mathcal{A}$, where $\mathcal{A}\equiv\mathcal{A}(m_{\eta}^2)$ is defined in terms of a loop integral, see Refs.~\cite{Masjuan:2015cjl,Sanchez-Puertas:2017sih} and references therein.\footnote{ In particular, and neglecting uncertainties, we take $\mathcal{A}(m_{\eta}^2)= -1.26 -5.47i$~\cite{Masjuan:2015cjl}.} The polarized decay yields\footnote{For that, we use the polarized spin projectors
    $u(p,\lambda\SLV{n})\bar{u}(p,\lambda\SLV{n}) = \frac{1}{2}\left(1+\lambda\gamma^5\slashed{n} \right)\left(\slashed{p}+m\right)$ and  
    $v(p,\lambda\SLV{n})\bar{v}(p,\lambda\SLV{n}) = \frac{1}{2}\left(1+\lambda\gamma^5\slashed{n} \right)\left(\slashed{p}-m\right)$. 
Particularly, $n^{\mu} = (0,\SLV{n}) \to (\gamma\beta n_z, n_T, \gamma n_z)$, with $|\SLV{n}|=1$.
 }
  \begin{multline}
    |\mathcal{M}(\lambda\SLV{n},\bar{\lambda}\bar{\SLV{n}})|^2 = \frac{m_{\eta}^2}{2} \Big[
      |g_P|^2 \left(1 - \lambda\bar{\lambda} [\SLV{n}\cdot\bar{\SLV{n}}] \right) 
      + |g_S|^2\beta_{\mu}^2 \big(1 - \lambda\bar{\lambda} [ n_z \bar{n}_z  -n_T\cdot\bar{n}_T ]   \big) \\
      +2\left[
         \lambda\bar{\lambda}\operatorname{Re}(g_Pg_S^*)(\bar{\SLV{n}}\times \SLV{n})\cdot \SLV{\beta}_{\mu} 
      + \operatorname{Im}(g_Pg_S^*)\SLV{\beta}_{\mu} \cdot(\lambda\SLV{n} -\bar{\lambda}\bar{\SLV{n}})
    \right] \Big], 
  \end{multline}
where---hereinafter---$n(\bar{n})$ will refer to the polarization axis for the $\mu^{+}(\mu^{-})$ (in its rest frame), the $\hat{z}$-axis will point along $\mu^+$ direction, and $\SLV{\beta}_{\mu}$ will refer to the $\mu^+$ velocity in the dimuon rest frame---here coinciding with the $\eta$ one. 
This would suffice to produce a MC for polarized decays.\footnote{In a real experiment the muon trajectory, its polarization, and subsequent---polarized---decay are accounted through {\textsc{Geant4}}~\cite{Agostinelli:2002hh}.} Still, in order to define the asymmetries and estimate their size  (in vacuum), we need to suplement this with the polarized muon decay (see \cref{sec:polmudec}), that leads to\footnote{We use, as it is standard, $\Gamma_{\gamma\gamma} = |F_{\eta\gamma\gamma}|^2 m_{\eta}^3\alpha^2\pi/4$ to normalize the result. Although we give $\tilde{g}_S^2$ terms for completeness, in the following we will only consider the interference with SM terms, which are $\tilde{g}_S$- rather than $\tilde{g}_S^2$-suppressed.} 
   \begin{multline}\label{eq:pllsm}
      \frac{d\Gamma}{\Gamma_{\gamma\gamma}} = 
      2\beta_{\mu} \left( \frac{\alpha m_{\mu}}{\pi m_{\eta}} \right)^2 \bigg[
        |\mathcal{A}|^2\left(1 + b\bar{b}\left\{ \SLV{\beta}\cdot\bar{\SLV{\beta}} \right\} \right)
        +2\beta_{\mu}\tilde{g_S} \big\{ b\bar{b} [(\SLV{\beta}\times\bar{\SLV{\beta}})\cdot\hat{z}]
                        \operatorname{Re}\mathcal{A} \\
       - (b\beta_z +\bar{b}\bar{\beta}_z) \operatorname{Im}\mathcal{A} \big\}
        +\tilde{g}_S^2\left( 1 +b\bar{b}\left\{ \beta_z\bar{\beta}_z 
                       -\SLV{\beta}_T\cdot\bar{\SLV{\beta}}_T \right\} \right)
     \bigg]  d_{e^{\pm}},
   \end{multline}
where $\tilde{g}_S = -g_S/(2m_{\mu}\alpha^2 F_{\eta\gamma\gamma})$ and $d_{e^{\pm}} = d\Omega dx d\bar{\Omega} d\bar{x} (4\pi)^{-2} n(x)n(\bar{x})$ refers to the $e^{+}e^{-}$ differential spectra, which is normalized to BR$(\mu\to e\nu\nu)\simeq 1$. The unbarred(barred) variables are kept for the $e^+(e^-)$ and $b\equiv b(x)$, with $n(x)$ and $b(x)$ defined below \cref{eq:mudecayapp}. If integrating over $d_{e^{\pm}}$, the terms in braces vanish, recovering the standard result for $\tilde{g}_S^2\to0$. 

For the \CPL{} scenario, $g_S = -\mathcal{C}$, and from \cref{eq:mathcalC}, $\tilde{g}_S = (0.510(c_{\ell equ}^{(1)2211} +c_{\ell edq}^{2211}) -0.771 c_{\ell edq}^{2222})$. For the \CPH{} one, the contribution is generated at the loop level and parallels the SM calculation. Defining $q(l)=p_{\mu^-} \pm p_{\mu^+}$, we find
   \begin{align}
    &  \tilde{g}_S = \frac{i F_{\eta\gamma\gamma}^{-1}}{\pi^2q^2\beta_{\mu}^2} \int d^4k \frac{F_{\eta\gamma^*\gamma^*}^{C\!P1}(k^2,(q-k)^2)w_1 + F_{\eta\gamma^*\gamma^*}^{C\!P2}(k^2,(q-k)^2)w_2}{k^2 (q-k)^2 ((p_{\mu^-}-k)^2-m_{\mu}^2)},  \\
   &   w_1 = k\cdot(q-k)l^2 -(k\cdot l)(k^2 +(q-k)^2), \quad 
      w_2 = k^2(q-k)^2\left( l^2 +2k\cdot l \right). \nonumber
   \end{align}
Using the form factors description in \cref{sec:TFF}, we find $\tilde{g}_S = (-0.87 -5.5i)\epsilon_1 + 0.66\epsilon_2$, where large hadronic uncertainties are implied.


In order to test our \CP{}-violating scenarios, we define the following asymmetries
   \begin{align}
     A_{L} \equiv&{} \frac{N(c_{\theta}>0) - N(c_{\theta}<0)}{N(\textrm{all})} = \bar{A}_{L} =
       \frac{\beta_{\mu}}{3} \frac{\operatorname{Im}\mathcal{A} \ \tilde{g}_S}{|\mathcal{A}|^2}, \\   
   A_{T} \equiv&{}  \frac{N( s_{\phi-\bar{\phi}}>0) - N( s_{\phi-\bar{\phi}}<0)}{N(\textrm{all})}  = 
       \frac{\pi\beta_{\mu}}{36} \frac{\operatorname{Re}\mathcal{A} \ \tilde{g}_S}{|\mathcal{A}|^2}, 
   \end{align}
where the barred version is the $A_{L}$ asymmetry for the $e^-$. As a result, we find
  \begin{align*}
      A_{L}^{H} &{}= 0.11 \epsilon_1 -0.04\epsilon_2, & 
      A_{L}^{L} &{}= -\operatorname{Im}(2.7(c_{\ell equ}^{(1)2211} +c_{\ell edq}^{2211}) -4.1c_{\ell edq}^{2222})\times 10^{-2}, \\
      A_{T}^{H} &{}= -0.07\epsilon_1 -0.002\epsilon_2, & 
      A_{T}^{L} &{}= -\operatorname{Im}(1.6(c_{\ell equ}^{(1)2211} +c_{\ell edq}^{2222}) -2.5 c_{\ell edq}^{2222})\times 10^{-3},
  \end{align*}
for the \CPH{} and \CPL{} scenarios, respectively. 
Taking BR($\eta\to\mu^+\mu^-)=5.8\times 10^{-6}$~\cite{Tanabashi:2018oca}, and the expected number of $\eta$ mesons at REDTOP ($2\times 10^{12}$)~\cite{Gatto:2016rae}, we obtain that the SM background for the asymmetry at the $1\sigma$ level is of the order of $N^{-1/2}=3\times 10^{-4}$. As a result, we find the following sensitivities: $\epsilon_{1(2)}\sim 10^{-3(2)}$ and $c_{\mathcal{O}}^{22st}\sim 10^{-2}$.

\subsection{The Dalitz decay: $\eta\to\gamma\mu^+\mu^-$}

\begin{figure}[t]\centering
  \includegraphics[width=0.4\textwidth]{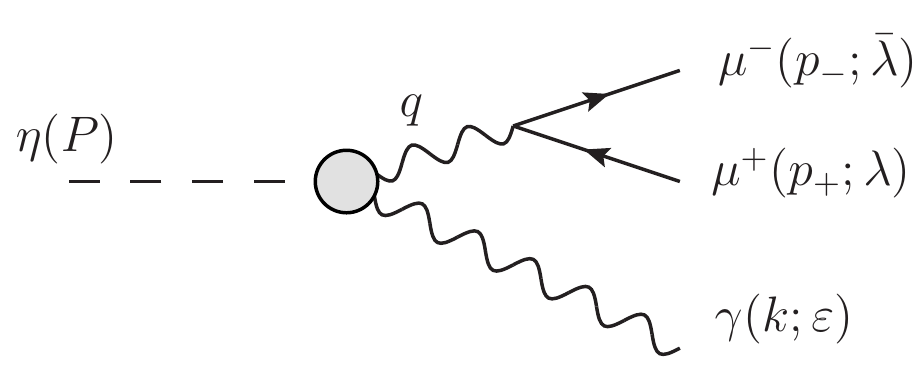} \qquad
  \includegraphics[width=0.22\textwidth]{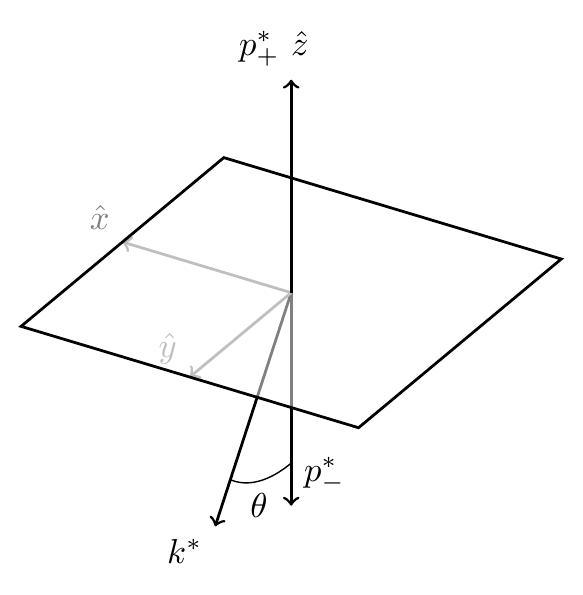} 
  \caption{Left: the LO SM contribution to the Dalitz decay. Right: the momentum labeling in the $\mu^+\mu^-$ reference frame.\label{fig:dalitz}}
\end{figure}
In the following, we introduce the---polarized---Dalitz decays: for simplicity we do not consider the most general amplitude, but the interference of the LO SM result with our \CP{}-violating amplitudes. Concerning the SM, the LO amplitude arises from the diagram in \cref{fig:dalitz} (left)\footnote{The $Z$ boson contribution is discussed in \cref{sec:Zboson} and does not affect the results here.}
  \begin{equation}\label{eq:dalitzsm}
    i\mathcal{M} = i e^3 \epsilon_{\mu\nu\rho\sigma}k^{\nu}q^{\sigma}\varepsilon^{\mu*} (\bar{u}\gamma^{\rho}v)F_{\eta\gamma\gamma^*}(q^2)q^{-2}.
  \end{equation}
Employing the phase space description in terms of the dilepton invariant mass $q^2= (p_+ + p_-)^2 \equiv s \equiv x_{\mu} m_{\eta}^2$ and the polar angle $\theta$ (see \cref{fig:dalitz} [right]), the differential decay width can be expressed as\footnote{We defined $y=\beta_{\mu}\cos{\theta}$ and $\beta_{\mu}^2=1-4m_{\mu}^2/s$.}
  \begin{equation}
    d\Gamma = \frac{1}{(2\pi)^3}\frac{m_{\eta}}{64}(1-x_{\mu})|\mathcal{M}|^2dx_{\mu} dy 
            = \Gamma_{\gamma\gamma} \frac{\alpha}{2\pi} \frac{1 -x_{\mu}}{m_{\eta}^2}\frac{|\mathcal{M}|^2}{|e^3F_{\eta\gamma\gamma}|^2}dx_{\mu} dy.
  \end{equation}
The LO SM result for the polarized Dalitz decay results in 
  \begin{multline}\label{eq:dalitzpolsm}
    |\mathcal{M}(\lambda\SLV{n},\bar{\lambda}\bar{\SLV{n}})|^2 = \frac{1}{4} \frac{e^6|F_{\eta\gamma\gamma^*}(s)|^2}{2s} (m_{\eta}^2 -s)^2 \Big[ 
      2  -\beta_{\mu}^2\sin^2\theta  + \lambda\bar{\lambda}\Big\{ \beta_{\mu}^2\sin^2\theta(n_z\bar{n}_z-n_T\cdot \bar{n}_T) \\
      +2\left( n_z\bar{n}_z\cos^2\theta +n_y\bar{n}_y\sin^2\theta  \right)
      -2\sqrt{1-\beta_{\mu}^2}\sin{\theta}\cos{\theta}(n_z\bar{n}_y +\bar{n}_z n_y) \Big\}
    \Big],
  \end{multline}
with similar conventions to those in the previous section (note that we choose, again, the $\mu^+$ to mark the $\hat{z}$ direction and the $\gamma$ to have an additional component along the $\hat{y}$ directions---see \cref{fig:dalitz} [right]). Once more, we include the muon decay to estimate the asymmetries, obtaining\footnote{$\tilde{F}_{\eta\gamma\gamma^*}(s)$ stands for the normalized transition form factor.} 
   \begin{multline}\label{eq:diffdalitzsm}
      \frac{d\Gamma}{\Gamma_{\gamma\gamma}} =  \frac{\alpha}{4\pi} \frac{|\tilde{F}_{\eta\gamma\gamma^*}(s)|^2}{s}(1-x_{\mu})^3 ds dy d_{e^{\pm}}
                \Big[
                     2  -\beta_{\mu}^2\sin^2\theta -b\bar{b}\Big\{
                     \beta_{\mu}^2\sin^2\theta(\beta_z\bar{\beta}_z -\beta_T\cdot \bar{\beta}_T) \\
                    +2\left( \beta_z\bar{\beta}_z\cos^2\theta +\beta_y\bar{\beta}_y\sin^2\theta  \right)
                    -2\sqrt{1-\beta_{\mu}^2}\sin{\theta}\cos{\theta}(\beta_z\bar{\beta}_y +\bar{\beta}_z \beta_y)
                \Big\}\Big].
   \end{multline}
Integrating over $d_{e^{\pm}}$, the terms in braces vanish and we obtain the standard result~\cite{Escribano:2015vjz,Husek:2017vmo}.
Concerning our \CP{}-violating scenarios, we give here the main results and relegate the intermediate steps to \cref{sec:ddres}. The final result reads
   \begin{align}
      \frac{d\Gamma_{C\!P_{\textrm{H}}}}{\Gamma_{\gamma\gamma}} = &
      \frac{\alpha}{\pi} \frac{\operatorname{Im}\tilde{F}_{\eta\gamma\gamma^*}(s)\tilde{F}_{\eta\gamma\gamma^*}^{C\!P1*}(s)}{s}(1-x_{\mu})^3 ds dy d_{e^{\pm}} \nonumber \\ & \quad\times
                \Big[
                     \sqrt{1-\beta_{\mu}^2}\sin{\theta}(b\beta_y -\bar{b}\bar{\beta}_y)
                     -\cos{\theta}(b\beta_z -\bar{b}\bar{\beta}_z)
                \Big],\\
      \frac{d\Gamma_{C\!P_{\textrm{HL}}}}{\Gamma_{\gamma\gamma}} = &
      \frac{\alpha}{\pi} \frac{(1- x_{\mu})}{s(1-y^2)} \frac{2\mathcal{C}ds dy d_{e^{\pm}} }{e^2 m_{\eta}F_{\eta\gamma\gamma}}  
               \Big[  \tilde{\alpha}_R\operatorname{Re}\tilde{F}_{\eta\gamma\gamma^*}(s) + \tilde{\alpha}_I\operatorname{Im}\tilde{F}_{\eta\gamma\gamma^*}(s)  \Big], 
   \end{align}
where $\tilde{\alpha}_{R,I}$ is obtained from the results in \cref{sec:ddres} upon $\alpha_{R,I}\to\alpha_{R,I}/m_{\eta}^3$ and, for $\alpha_{R}$, $\lambda n(\bar{\lambda}\bar{n}) \to b\beta(\bar{b}\bar{\beta})$ while, for $\tilde{\alpha}_I$, $\lambda n(\bar{\lambda}\bar{n}) \to -b\beta(+\bar{b}\bar{\beta})$. 
In the following, we introduce two additional asymmetries besides those defined in \cref{sec:dilepton},
   \begin{equation}
     A_{L\gamma} \equiv \frac{N(s_{\phi}>0) - N(s_{\phi}<0)}{N(\textrm{all})}, \quad
     A_{TL} \equiv \frac{ N(c_{\phi}c_{\bar{\theta}}>0) - N(c_{\phi}c_{\bar{\theta}}<0) }{ N(\textrm{all}) }.
   \end{equation}
While for the SM result [\cref{eq:dalitzpolsm}] these asymmetries vanish, in our \CP{}-violating scenarios we find\footnote{For analytic results in terms of phase-space integrals, see \cref{sec:ddres}. In these results, we use the form factors in \cref{sec:TFF} and assume the \CP{}-violating form factors to be real.}
   \begin{align}
   A_{L}^{H} &{}= 0 &    A_{L}^{HL} &{}= -4\operatorname{Im}(1.1(c_{\ell equ}^{(1)221} +c_{\ell edq}^{221}) -1.7c_{\ell edq}^{2222})\times 10^{-7}, \\ 
   A_{L\gamma}^{H} &{}= -0.002\epsilon_1 & A_{L\gamma}^{HL} &{}= 5\operatorname{Im}(1.1(c_{\ell equ}^{(1)2211} +c_{\ell edq}^{2211}) -1.7c_{\ell edq}^{2222})\times 10^{-6}, \\
   A_{TL}^{H} &{}= 0 &    A_{TL}^{HL} &{}= 2\operatorname{Im}(1.1(c_{\ell equ}^{(1)221} +c_{\ell edq}^{221}) -1.7c_{\ell edq}^{2222})\times 10^{-5}, \\ 
   A_{T}^{H} &{}= 0 &    A_{T}^{HL} &{}= -5\operatorname{Im}(1.1(c_{\ell equ}^{(1)221} +c_{\ell edq}^{221}) -1.7c_{\ell edq}^{2222})\times 10^{-6} . 
   \end{align}
Taking BR($\eta\to\mu^+\mu^-\gamma)=3.1\times 10^{-4}$~\cite{Tanabashi:2018oca}, we obtain that the SM background for the asymmetry at the a $1\sigma$ level is of the order of $10^{-5}$ for REDTOP statistics, which results in the following sensitivities: $\epsilon_1\sim 10^{-2}$ and $c_{\mathcal{O}}^{22st}\sim1$. For completeness, we show in \cref{sec:Zboson} that the $Z$-boson parity-violating asymmetry is irrelevant for such statistics.

\subsection{Classical channel: $\eta\to\mu^+\mu^-e^+e^-$}

The double Dalitz decay has been the standard way to test for \CP{}-violation in pseudoscalar mesons decays, as it does not require to measure the polarization of the leptons~\cite{Samios:1962zza,Abouzaid:2008cd,Uy:1990hu}. In this study, we restrict ourselves to the $\eta\to\mu^+\mu^-e^+e^-$ decay\footnote{The SMEFT operators involving electrons are tightly constrained as we shall see and we neglect them, while the purely muonic channel is less interesting since its BR is two orders of magnitude below~\cite{Kampf:2018wau}.} and study the interference terms alone. Concerning the SM result, notation, etc., we refer to Ref.~\cite{Kampf:2018wau}.
Regarding the \CPH{} interaction, we recover the results in \cite{Kampf:2018wau} with the addition of the the second form factor that was omitted there,
  \begin{multline}\label{eq:ddcp1}
    \frac{d\Gamma_{C\!P_{\textrm{H}}}}{\Gamma_{\gamma\gamma}} = \frac{\alpha^2}{32\pi^3} \operatorname{Re}
    \frac{\tilde{F}_{\eta\gamma^*\gamma^*}}{s_{12}s_{34}} \lambda^2\Big[  
     y_{12}y_{34}\sqrt{w^2(\lambda_{12}^2-y_{12}^2)(\lambda_{34}^2-y_{34}^2)}\Big( 2\tilde{F}_{\eta\gamma^*\gamma^*}^{C\!P1*} 
         -z m_{\eta}^2\tilde{F}_{\eta\gamma^*\gamma^*}^{C\!P2*} \Big) \\
     - (\lambda_{12}^2-y_{12}^2)(\lambda_{34}^2-y_{34}^2)\cos\phi\left(2z\tilde{F}_{\eta\gamma^*\gamma^*}^{C\!P1*} -w^2m_{\eta}^2\tilde{F}_{\eta\gamma^*\gamma^*}^{C\!P2*} \right) 
    \Big]\sin\phi \ d\Phi.
  \end{multline}
Concerning the \CPL{} scenario, there are four different contributions. Those arising from the effective operators coupling to muons are
   \begin{align}
      i\mathcal{M}_1 &{}= -i\frac{e^2 \mathcal{C}}{s_{34}(p_{134}^2 -m_a^2)}
                 \big[ \bar{u}_1\gamma^{\mu}(\slashed{p}_{134} +m_a)v_2\big] \big[\bar{u}_3\gamma_{\mu}v_4 \big], \\                             
      i\mathcal{M}_2 &{}= i\frac{e^2 \mathcal{C}}{s_{34}(p_{234}^2 -m_a^2)}
                 \big[ \bar{u}_1\gamma^{\mu}(\slashed{p}_{234} -m_a)v_2\big] \big[\bar{u}_3\gamma_{\mu}v_4 \big],                             
   \end{align}
while, if considering the coupling to electrons, the remaining two would be obtained upon $1(2)\to 3(4)$ and $\mu\to e$ exchange. Their contribution to the differential decay width reads
   \begin{multline}\label{eq:ddcp2}
       \frac{d\Gamma_{C\!P_{\textrm{HL}}}}{\Gamma_{\gamma\gamma}} =
      \frac{\alpha}{16\pi^4}\frac{\operatorname{Re}\tilde{F}_{\eta\gamma^*\gamma^*} }{s_{12}s_{34}}\lambda^2\sin\phi
       \Bigg[
       \frac{m_{\mu}\mathcal{C}}{F_{\eta\gamma\gamma}}\frac{2x_{34}y_{12}y_{34}(1+\delta)  -(1-\delta)\Xi}{s_{34}[(1-\delta)^2 -\lambda^2 y_{12}^2]} \Bigg]
  \\ 
       \times\sqrt{w^2(\lambda_{12}^2-y_{12}^2)(\lambda_{34}^2-y_{34}^2)} \ d\Phi.
   \end{multline}
As said, in these decays a polarization analysis is not required to test for \CP{} violation; this is related to the lepton plane  angular asymmetries. Defining
   \begin{equation}
      A_{\phi/2} = \frac{N(s_{\phi}c_{\phi}>0) - N(s_{\phi}c_{\phi}<0)}{N(\textrm{all})},
   \end{equation}  
we obtain
  \begin{align}
    A_{\phi/2}^H &{}= -\frac{\alpha^2}{9\pi^3}\frac{1}{N} \int \operatorname{Re}
      \frac{\tilde{F}_{P\gamma^*\gamma^*}}{s_{12}s_{34}} \lambda^2\lambda_{12}^3\lambda_{34}^3 
      \left(2z\tilde{F}_{\eta\gamma^*\gamma^*}^{C\!P1*} -w^2m_{\eta}^2\tilde{F}_{\eta\gamma^*\gamma^*}^{C\!P2*} \right)   ds_{12}ds_{34}, \\
     A_{\phi/2}^{HL} &{}= -\frac{4\alpha}{3\pi^4} \frac{1}{N}
                  \int \frac{\operatorname{Re}\tilde{F}_{\eta\gamma^*\gamma^*}}{\lambda m_{\eta}^4F_{\eta\gamma\gamma}} 
           \frac{m_{\mu}\mathcal{C}}{s_{34}} \lambda^3_{34} 
               \bigg(  \lambda\lambda_{12}(1-\delta) 
           - [(1-\delta)^2 -\lambda^2\lambda^2_{12}]\tanh^{-1}\left[\frac{\lambda\lambda_{12}}{1\mp\delta}\right] \bigg)
           ds_{12}ds_{34}.
  \end{align}
Employing the form factors defined in \cref{sec:TFF}, we obtain
   \begin{equation}
      A_{\phi/2}^{H} = -0.2\epsilon_1 +0.0003\epsilon_2, \quad
      A_{\phi/2}^{HL} = -\operatorname{Im}(1.3(c_{\ell e qu}^{(1)2211} +c_{\ell edq}^{2211}) -1.9c_{\ell edq}^{2222})\times 10^{-5}. 
   \end{equation}
Taking  BR$(\eta\to\mu^+\mu^-e^+e^-)=2.3\times 10^{-6}$~\cite{Kampf:2018wau}, we find the SM background at REDTOP at the 1$\sigma$ level to be $5\times 10^{-4}$. Consequently, we are sensitive to $\epsilon_1\sim 10^{-3}$ and $c_{\mathcal{O}}\sim40$.

\section{Bounds from neutron dipole moment}\label{sec:nedm}

The interaction of a charged fermion with the electromagnetic current $(j^{\mu})$ can be expressed as $\bra{\ell(p')} j^{\mu} \ket{\ell(p)} =\mathcal{Q}_{\ell} \bar{u}_{p'} \Gamma^{\mu}(q) u_p$, where~\cite{Czarnecki:1900zz}\footnote{In general, $F_1(0)=1$, except for a neutral fermion, such the neutron, where we take $\mathcal{Q}_n=1$ and $F_1(0)\equiv 0$.} 
   \begin{equation}\label{eq:ffs}
     \Gamma^{\mu} =  \gamma^{\mu} F_1(q^2) +\frac{i\sigma^{\mu\nu}q_{\nu}}{2m_{\ell}}F_2(q^2)
                          -\frac{\sigma^{\mu\nu}q_{\nu}}{2m_{\ell}}\gamma^5 F_E(q^2) +(q^2\gamma^{\mu} -\slashed{q}q^{\mu})\gamma^5 F_{A}(q^2),
   \end{equation}
with $q(l)=p'\mp p$. At low energies, $F_2$ and $F_E$ generate magnetic and electric dipole moments, respectively. Particularly, in their non-relativistic limit\footnote{\label{fn:dipole}Usually $\mu$ is given in units of $e\hbar/2m_\ell$ and $F_2(0)$ yields the anomalous magnetic moment. The electric dipole moment commonly refers to $d$ in units of $e\mskip3mu\textrm{cm}$, such that it involves $(\hbar c [\textrm{GeV cm}])/(2m_{\ell}c^2 [\textrm{GeV}])F_E(0)$. Also, we take $\mathcal{L} = \bar{\psi}(i\gamma^{\mu}D_{\mu} -m)\psi$ with $D_{\mu} = \partial_{\mu} -ie\mathcal{Q}_{\ell}A_{\mu}$ so that $i\mathcal{M} = ie \bar{u}_{p'} \Gamma^{\mu} u_p \epsilon_{\mu}$. With these definitions, the dipole moments can be also obtained from the effective Lagrangian $\mathcal{L} = \mathcal{Q}_{\ell}\frac{e}{2}\bar{\psi}\sigma^{\mu\nu}(\mu + i\gamma^5d)\psi F_{\mu\nu}$.} 
  \begin{equation}
    e\Gamma^{\mu}A^{\textrm{cl}}_{\mu} \stackrel{\text{NR}}{=} -\mu \SLV{\sigma}\cdot\SLV{B} -d \SLV{\sigma}\cdot\SLV{E}, 
    \qquad \mu = \frac{e\hbar}{2m_\ell}(F_1(0)+F_2(0)), \quad	d = \frac{e\hbar}{2m_\ell c}F_E(0).
  \end{equation}
Being suppressed in the SM, EDMs put severe constraints on \CP{}-violating new physics scenarios~\cite{Pruna:2017tif}. 
In addition, the dipole moments of heavy atoms and molecules put strong constraints for contact \CP{}-violating electron-quark $D=6$ operators~\cite{Yanase:2018qqq}. This is the reason for which we did not consider the electronic, but the muonic case---see also in this respect the implications of the  recent ACME Coll.~\cite{Andreev:2018ayy} results for the  electron EDM in Ref.~\cite{Cesarotti:2018huy}.
In the sections below, it will be useful to employ projectors (in analogy to Refs.~\cite{Barbieri:1972as} for the magnetic moment) for $F_E$ which, in $D=4$ dimensions read\footnote{$\Gamma^{\mu} = -q_{\rho}\Gamma^{\mu\rho}$, where $\Gamma^{\mu\rho}\equiv \lim_{q\to 0} \partial_{q_{\mu}}\Gamma^{\rho}$.}
  \begin{equation}
    F_E(q^2) = \operatorname{tr} \frac{-im l^{\mu} \gamma^5}{q^2(q^2-4m^2)}
                                  (\slashed{p}'+m)\Gamma^{\mu}(\slashed{p}+m), \qquad
    F_E(0) = \frac{i}{12m^2}p_{\mu} \operatorname{tr} (\slashed{p}+m) \gamma_{\rho}\gamma^5 (\slashed{p}+m)\Gamma^{\mu\rho}.
  \end{equation}
In the following, we discuss the bounds that the nEDM puts on our new physics scenarios, for which we employ the projector in the $q\to 0$ limit, in which dipole moments are defined.

\subsection{nEDM bounds on \CPH{} scenario}\label{sec:nedmh}

\begin{figure}\centering
   \includegraphics[width=0.6\textwidth]{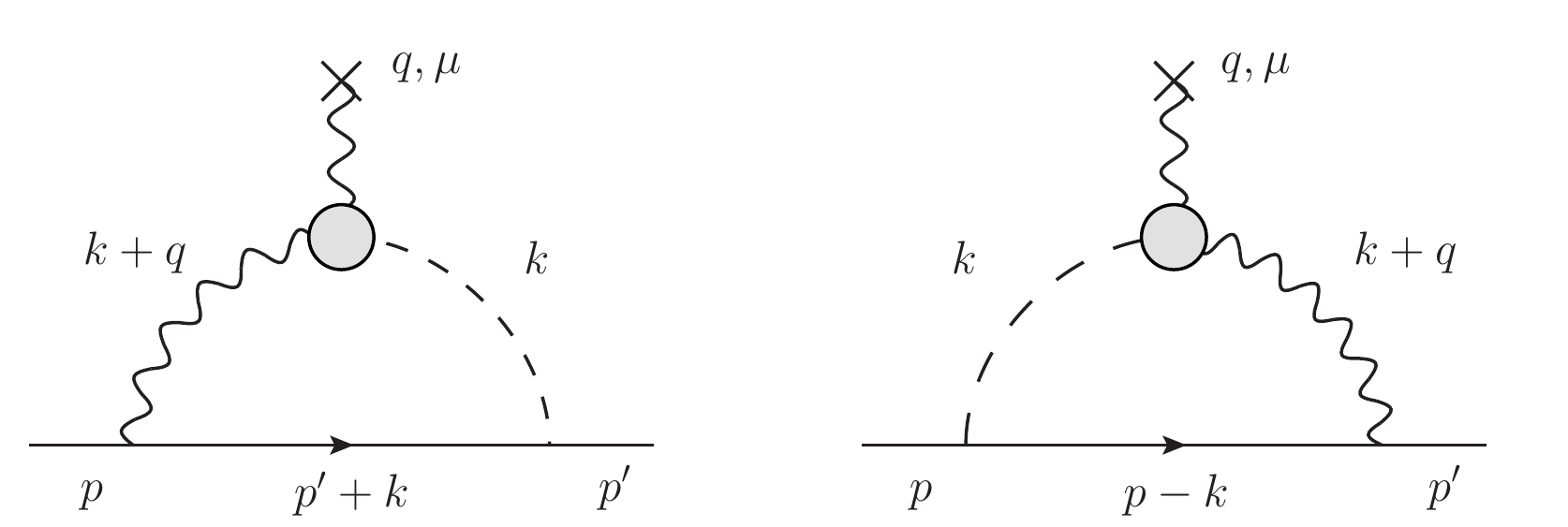}
   \caption{Contributions to the nucleon EDM via a \CP-violating $\eta$ coupling to $\gamma\gamma$.\label{fig:edm1}}
\end{figure}

As stated in \cref{sec:cpscen}, there are a number of effective operators belonging to this case---each of them contributing differently to the nEDM and posing an individual challenge. However, for our purpose---as we shall see---it will suffice to account that the \CP{}-violating $\eta$ TFF will generate a nEDM via the diagrams in \cref{fig:edm1}, which amplitudes read\footnote{For the $NN\eta$ coupling we take the results in Ref.~\cite{Gutsche:2016jap}, where this was given by $\mathcal{L} \supset \frac{g_{\eta NN}}{2F_{\eta}} \bar{N}\gamma^{\mu}\gamma^5N\partial_{\mu}\eta$ with $g_{\eta NN}=0.673$ and $F_{\eta}=1.37 F_{\pi}$.}
   \begin{align}
      i\mathcal{M}_1 &{}= ie\frac{-e^2g_{\eta NN}}{2F_{\eta}} \int \frac{d^4k}{(2\pi)^4}
                     \frac{\bar{u}_{p'} \Gamma_{\nu}(-k) (\slashed{p}'+\slashed{k} +m_N) (\slashed{k}+\slashed{q})\gamma^5 u_p}
                          {k^2[(k+q)^2-m_{\eta}^2][(p'+k)^2-m_{N}^2]}, \\
      i\mathcal{M}_2 &{}= ie\frac{-e^2g_{\eta NN}}{2F_{\eta}} \int \frac{d^4k}{(2\pi)^4}
                     \frac{\bar{u}_{p'} (\slashed{k}+\slashed{q})\gamma^5 (\slashed{p} -\slashed{k} +m_N) \Gamma_{\nu}(-k) u_p}
                          {k^2[(k+q)^2-m_{\eta}^2][(p-k)^2-m_{N}^2]}.
   \end{align}
Regarding the $\Gamma_{\nu}$ vertex, we take it to be given by the on-shell form factors $F_{1,2}$ in \cref{eq:ffs}, which closely follows the methodology in Ref.~\cite{Gutsche:2016jap}. Of course, this contribution is rather model dependent and there will be additional ones, but should be enough to provide an order-of-magnitude estimate. Using the projector technique, we obtain
   \begin{align}
     F_E(0)={} & \frac{g_{\eta NN}}{6F_P}\frac{\alpha}{\pi} \frac{16\pi^2}{i} \int \frac{d^4 k}{2\pi^4} 
                F_{\eta\gamma^*\gamma^*}^{C\!P1}(k^2,0)\Bigg[
                \frac{k^2[2m_N^2 -3(k\cdot p)] -2(k \cdot p)^2}
                     {k^2(k^2 -m_{\eta}^2)( (p+k)^2 -m_N^2 ) } F_1(k^2) \nonumber\\ & \qquad
         +      \frac{ 4(k\cdot p)^3 +2k^2(k\cdot p)[(k\cdot p) -2m_N^2] +m_N^2 k^4}
                     {k^2(k^2 -m_{\eta}^2)( (p+k)^2 -m_N^2 ) } \frac{F_2(k^2)}{2m_N^2}
         \Bigg], \\
           ={} & \epsilon_1 F_{\eta\gamma\gamma}
           \frac{g_{\eta NN}}{6F_{\eta}} \frac{\alpha}{\pi} \int_0^{\infty}dK^2\frac{K^2}{K^2+m_{\eta^2}}
                 \tilde{F}_{\eta\gamma^*\gamma^*}^{C\!P1}(-K^2,0)(1-\beta) \nonumber\\ & \qquad 
          \times\left(
                  F_2(-K^2)\frac{3K^2}{16m_N^2}(3 -\beta)  -F_1(-K^2)\left[ 1 + (1+\beta)^{-1} \right]
               \right), \label{eq:edmcph}
   \end{align}
where $\beta = (1+\frac{4m_N^2}{K^2})^{1/2}$ and in the second line we have used the Gegenbauer polynomials technique~\cite{Knecht:2001qf}. For the numerical evaluation, we employ the TFFs description in \cref{sec:TFF} and the eletromagnetic form factors parametrization in Ref.~\cite{Kelly:2004hm}, obtaining $d_E^p = 9.6\times 10^{-19} \epsilon_1$ and $d_E^n = -6.2\times 10^{-20} \epsilon_1$, in units of $e\mskip3mu\textrm{cm}$. Accounting that $d_E^{p(n)}= 2.1(0.30)\times 10^{-25}$, this places bounds which are orders of magnitude beyond the experimental sensitivities accessible at REDTOP.
Of course, this offers no bounds on $\epsilon_2$ and an alternative would be to set $F_{\eta\gamma\gamma}^{C\!P1,2}(0,0)\to 0$---this is, a vanishing coupling to real photons. A tuning such that would be suspicious however without a dynamical origin and we conclude that \CP{}-violating physics in the context of \CP{}-violating $\eta\gamma^*\gamma^*$ interactions are out of reach for any experiment so far.

\subsection{nEDM bounds on \CPL{} scenario}

\begin{figure}\centering
   \includegraphics[width=\textwidth]{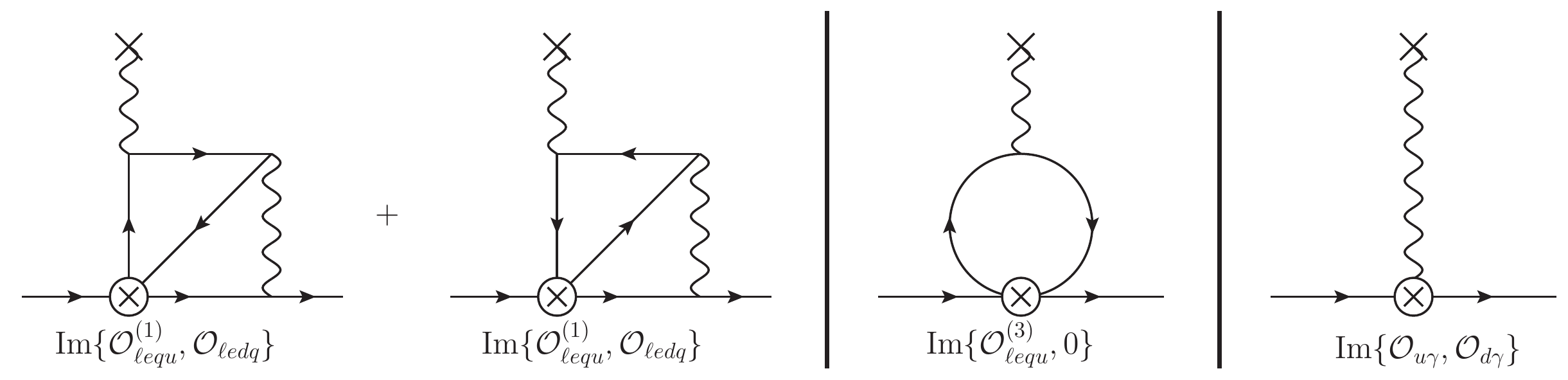}
   \caption{\CPL{} scenario contribution to the nEDM: first contribution at two loops (left) (additional reversed diagrams appear). The $\mathcal{O}_{lequ}^{(1)}$ operator requires the $\mathcal{O}_{lequ}^{(3)}$ operator for renormalization at the one loop level (center)---which requires a dipole counterterm at one-loop too (right). Dipole operators also enter when renormalizing the $\mathcal{O}_{ledq}$ operator at two loops.\label{fig:cp2edm}}
\end{figure}

Here, there is no mechanism inducing a dipole moment at one loop, which can be related to the fact that the Green's function $\bra{0} T\{ V^{\mu}(x) S(P)(0) \} \ket{0}$ vanishes in QED+QCD due to charge conjugation. The first contribution appears at two-loops and requires renormalization, which is sketched at the quark level in \cref{fig:cp2edm}. This involves the following operators
  \begin{align}
    \mathcal{O}_{\ell equ}^{(3)} =&{} \ ... \
        \to-i\frac{\operatorname{Im}c_{\ell equ}^{(3)prst}}{2v^2}
               \left[ (\bar{e}_p\sigma^{\mu\nu}\gamma^5e_r)(\bar{u}_s\sigma_{\mu\nu}u_t) +(\bar{e}_p\sigma^{\mu\nu}e_r)(\bar{u}_s\sigma_{\mu\nu}\gamma^5u_t) \right], \\
    \mathcal{O}_{qB(W)} =&{} \ ... \ \to
  i\frac{\operatorname{Im}c_{q\gamma}^{st}}{v} \left(\bar{q}_s \sigma^{\mu\nu} \gamma^5  q_t \right) F_{\mu\nu}
 \quad c_{u(d)\gamma}^{st}=\frac{c_{u(d)B}^{st}c_w\pm c_{u(d)W}^{st}s_w}{\sqrt{2}}.
  \end{align}
For the nucleon, the \CP{}-violating contribution to the electromagnetic vertex is
 \begin{align}
    \bar{u}_{p'}\Gamma^{\mu}u_p &{}= e^2\sum_i \int \frac{d^4 k}{(2\pi)^4} \frac{1}{k^2} 
                   \bigg[\frac{1}{i}\int e^{i k\cdot z} \bra{N_{p'}}
                       T\{  j_{\nu}(z) (\bar{q}\Gamma_i q)(0) \} \ket{N_p} \bigg] \nonumber\\
                 &{} \qquad\qquad \  \times\bigg[\frac{1}{i}\int e^{-i (q\cdot x +k\cdot y)} 
                       \bra{0}T\{ j^{\mu}(x)j^{\nu}(y)(\bar{\ell}\tilde{\Gamma}_i \ell)(0) \} \ket{0} \bigg] \nonumber\\
                &{}\equiv e^2\sum_i \int \frac{d^4 k}{(2\pi)^4} \frac{1}{k^2} 
                      \Pi_{NNV\Gamma_i}^{\rho}(-k,k+q) \Pi_{VV\tilde{\Gamma}_i}^{\mu\nu}(k,q) g_{\nu\rho},
 \end{align}
where, for $\mathcal{O}_{\ell equ(\ell edq)}$, we have the combinations $\sum_i\{\Gamma_i,\tilde{\Gamma}_i\} = -\frac{c_{\ell equ(\ell edq)}}{2v^2} [ \{ i\gamma^5, 1  \} \pm \{ i\gamma^5, 1  \} ] $.\footnote{Note in particular that potential additional diagrams with a single photon attached to the lepton line will be related again to $\bra{0}T\{V^{\mu}(x) S(P)(0)\}\ket{0}=0$.} 
  \begin{figure}\centering
     \includegraphics[width=\textwidth]{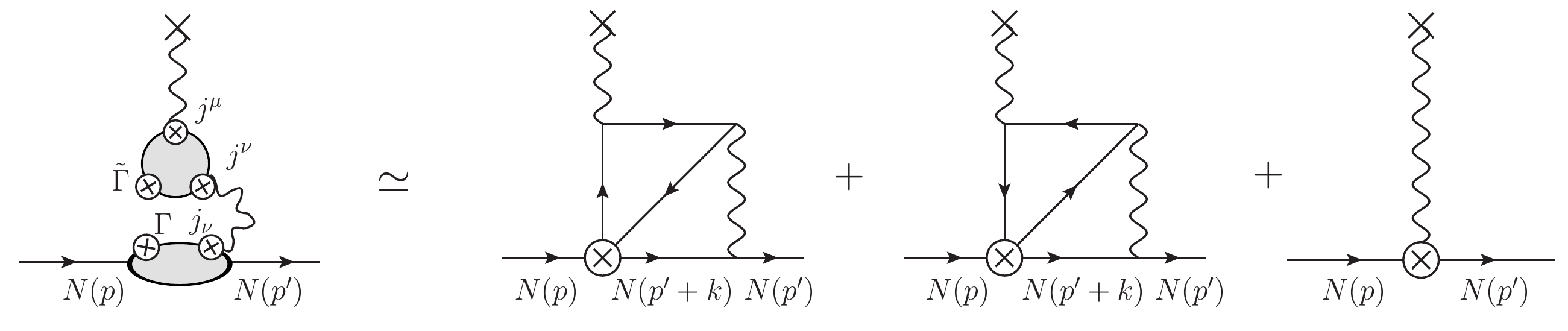}
     \caption{The generic contribution to a nucleon (N) EDM (additional counterterms need to be included as well). We approximate it as a low-energy (finite) contribution saturated via an intermediate nucleon (N) state (reversed diagrams implied) and a high-energy contribution (including counterterms) that mimics a contact term resulting from the OPE of the two currents. \label{fig:EDMn}}
  \end{figure}
In the following, we will simplify the calculation to get an order-of-magnitude estimate as follows: in the low-energy region (which we take below 2~GeV), we will assume the hadronic blob to be dominated by an intermediate neutron state, as shown in \cref{fig:EDMn}. Above, we employ the operator product expansion (OPE) for large (euclidean) momentum $k$.

Concerning the low-energy part, we have two hadronic elements to be computed. For $\Gamma=i\gamma^5$, we approximate such an interaction via an intermediate pseudo-Goldstone boson state ($\pi^0,\eta,\eta'$)---similar to the \CPH{} scenario. Regarding $\Gamma=1$, we approximate it via the scalar form factor (see \cref{app:scalarff}). For the electromagnetic form factors we use again Ref.~\cite{Kelly:2004hm}.
Regarding $\Pi^{\mu\nu}_{VVS(P)}(k,q)$, we provide them in the vanishing $q\to 0$ limit
  \begin{align}
     \Pi_{VVP}^{\mu\nu}(0,k) &{}= \epsilon^{\nu\mu q k}
                               \frac{-8m_{\ell}}{K^2\beta_{\ell}}
                               \ln\left( \frac{\beta_{\ell} +1}{\beta_{\ell} -1} \right), \\
     \Pi_{VVS}^{\mu\nu}(0,k) &{}= (g^{\mu\nu}(k\cdot q) -k^{\mu}q^{\nu}) 
 \frac{i}{16\pi^2}\frac{8m_{\ell}}{K^2\beta_{\ell}}\left[
       \frac{1 +\beta_{\ell}^2}{2\beta_{\ell}}\ln\left( \frac{\beta_{\ell} +1}{\beta_{\ell} -1} \right) -1
     \right],
  \end{align}
up to $\mathcal{O}(q^2)$ corrections, where $\beta_{\ell}^2=1+4m_{\ell}^2K^{-2}$ and $K^2=-k^2$. We obtain\footnote{From Ref.~\cite{Gutsche:2016jap}, we have $g_{\pi NN}=g_A=1.27$, $g_{\eta(\eta') NN}=0.67(1.17)$ and $F_{\eta(\eta')}=1.37(1.16)F_{\pi}$ with $F_{\pi}=92$~MeV. Concerning the pseudoscalar matrix elements, and following Ref.~\cite{Feldmann:1999uf}, $h_{\pi}^u=-h_{\pi}^d=F_{\pi}m_{\pi^0}^2 \hat{m}^{-1}=0.48~\textrm{GeV}^2$, $h_{\eta(\eta')}^{u,d}=F_{\eta(\eta)'}^q m_{\pi^0}^2 \hat{m}^{-1}=0.40(0.35)~\textrm{GeV}^2$ and $h_{\eta(\eta')}^{s}=F_{\eta(\eta)'}^s(2m_K^2 -m_{\pi^0}^2) m_s^{-1}=-0.42(0.53)~\textrm{GeV}^2$.}
  \begin{align}
     F_E^{N;q}(0) = \operatorname{Im}c_{\ell equ(dq)}\frac{\alpha}{\pi} \frac{G_F m_{\ell}}{6\sqrt{2}\pi^2} 
                \int_0^{\infty} dK \left[ m_N H_S \pm \frac{g_{PNN}h_P^q}{F_P} H_P \right],
  \end{align}
with $h_P^q =\bra{0} \bar{q} i\gamma^5 q \ket{P}$, $q=\{u,d,s\}$ a flavor index, and the functions
  \begin{align}
    H_S(K)&{}=\frac{K}{m_N^2\beta_{\ell}}\left(\frac{2\beta_N^2}{1+\beta_N} -1\right)
       \ln\left( \frac{\beta_{\ell}+1}{\beta_{\ell}+1} \right) F_S^{N;q}(-K^2) \left[ F_1^N(-K^2) +F_2^N(-K^2) \right], \\
    H_P(K)&{}=\frac{K \beta_{\ell}^{-1}}{K^2 +m_P^2}
       \left(\frac{1+\beta_{\ell}^2}{2\beta_{\ell}}  
         \ln\left( \frac{\beta_{\ell}+1}{\beta_{\ell}+1} \right) -1 \right) \nonumber
       \bigg[ F_1^N(-K^2)[1+ (1+\beta_N)^{-1}] \\ &{}\qquad\qquad\qquad\qquad\qquad\qquad\qquad\qquad\qquad\qquad\qquad +F_2^N(-K^2)\frac{3K^2}{16m_N^2}[3-\beta_N] \bigg].
  \end{align}
Numerically we find\footnote{We checked that the integral saturates at 2~GeV. The errors for the neutron are shown to illustrate the impact on the scalar form factor model only---dominated by the $\sigma_{\pi N}$ term.}
  \begin{align}
    d_E^p &{}= \operatorname{Im}
       (2.55c_{\ell equ}^{(1)2211} -3.17c_{\ell edq}^{2211} -0.18c_{\ell edq}^{2222})\times10^{-23}, \\
    d_E^n &{}= \operatorname{Im}
       (-0.75(26)c_{\ell equ}^{(1)2211} +0.92(27)c_{\ell edq}^{2211} +0.08(1)c_{\ell edq}^{2222})\times10^{-23}.\label{eq:led}
  \end{align}

For the high-energy region, the OPE calculation for the two-currents (to be included in the final hadronic matrix element) parallels that at the quark level---the main difference being the scale. Assuming that the theory was renormalized at a scale close to the electroweak one and assuming the quark dipole moment negligible at such scale, the result can be estimated by the large logs. To find these, we opt to use a cutoff regularization $(\Lambda)$ for the quark diagram level leading to 
  \begin{align}
    F_E^q(0)={}&\frac{\alpha}{\pi}\frac{G_Fm_{\ell}m_q}{6\sqrt{2}\pi^2}\mathcal{Q}_q\int_0^{\infty} 
                 \frac{dK K}{m_q^2\beta_{\ell}}\bigg[ 
                     \left(\beta_q -\frac{2}{1+\beta_q}\right)\left[ 
                       \frac{1+\beta_{\ell}^2}{2\beta_{\ell}}
                       \ln\left(\frac{\beta_{\ell}+1}{\beta_{\ell}-1}\right) 
                       -1\right]
       \nonumber \\  &           \pm
                  \left(\frac{2\beta_q^2}{1+\beta_q} -1\right)  
                     \ln\left(\frac{\beta_{\ell}+1}{\beta_{\ell}-1}\right) 
                \bigg]\operatorname{Im}c_{\ell equ(dq)} \\ &
       \to \frac{\alpha}{\pi} \frac{G_F m_{\ell} m_q}{6\sqrt{2}\pi^2}
           \left\{ \operatorname{Im}c_{\ell equ}^{(1)}(\ln^2\Lambda^2 -\ln\Lambda^2) , 
                    \operatorname{Im}c_{\ell edq}\ln\Lambda\right\}.
  \end{align}
As a check, the $\ln^2\Lambda$ terms reproduce the expectation from the one-loop RG equations.\footnote{With our conventions, $\frac{d c_{\ell equ}^{(3)}}{d \ln\mu} \supset -\mathcal{Q}_e\mathcal{Q}_u \frac{\alpha}{2\pi} c_{\ell equ}^{(1)}$ and $\frac{d c_{u\gamma}}{d \ln\mu} \supset - e\frac{m_{\mu}\mathcal{Q}_e}{2\pi^2v}  c_{\ell equ}^{(3)}$---in agreement with Ref.~\cite{Jenkins:2017dyc}.} Moreover, we find good agreement for the $\ln\Lambda$ term (which represents the leading log for $\mathcal{O}_{\ell edq}$) comparing to the recent results in Ref.~\cite{Panico:2018hal}.\footnote{In particular, with Eq.~(2.35) in \cite{Panico:2018hal} one takes $e\leftrightarrow d$ and $N_c=1$. One also needs to take care of the sign conventions---which are essentially related to our opposite choice for the covariant derivative.} From the neutron matrix elements $g_T^q \equiv \bra{n} \bar{q}\sigma^{\mu\nu}\gamma^5q \ket{n}$, obtained from lattice QCD at $\mu=2$~GeV~\cite{Bhattacharya:2015esa}, and using the renormalization scale $\mu_0=100$~GeV, we obtain for the high-energy contribution
  \begin{equation}\label{eq:hed}
        d_E^n = \operatorname{Im}
       (-0.59c_{\ell equ}^{(1)2211} +0.15c_{\ell edq}^{2211} +0.001c_{\ell edq}^{2222})\times10^{-23},
  \end{equation}
which is subleading compared to the low-energy contribution. Adding up \cref{eq:led,eq:hed} and assuming uncorrelated Wilson coefficients, we find that the nEDM puts the following constraints
  \begin{equation}
    \operatorname{Im}c_{\ell equ}^{(1)2211} < 0.002, \qquad
    \operatorname{Im}c_{\ell edq}^{2211} < 0.003, \qquad
    \operatorname{Im}c_{\ell edq}^{2222} < 0.04. \qquad
  \end{equation}
Once more, we emphasize that large uncertainties are implied, thus these should be taken as an order-of-magnitude estimate. As a conclusion, we find that $\eta\to\mu^+\mu^-$ decays are the only ones that might show \CP{}-violating signatures for $c_{\ell edq}^{2222}\simeq 10^{-2}$.

\section{Conclusions and Outlook}\label{sec:conc}

In this study, we have examined different imprints of \CP{} violation arising from the SMEFT in different $\eta$ muonic decays, which are effectively encoded via \CP{}-violating transition form factors or contact $\eta$-lepton interactions. Having in mind the REDTOP experiment---a proposed $\eta$ factory with the ability to measure the polarization of muons---we have estimated the sensitivities that can be reached in each case. After computing the implications of these scenarios on the nEDM, we have found that only $\eta$-lepton interactions---particularly the $\mathcal{O}_{\ell edq}^{2222}$ operator---might leave an imprint via the muons polarization in the $\eta\to\mu^+\mu^-$ decay.\footnote{Being this a potential channel to look for \CP{} violation, one might wonder about its $\eta'$ counterpart. An analogous computation shows $A_{L}^{L} = -\operatorname{Im}(1.4(c_{\ell equ}^{(1)2211} +c_{\ell edq}^{2211}) +2.9c_{\ell edq}^{2222})\times 10^{-2}$. Since BR$(\eta'\to\mu^+\mu^-)\simeq 1.4\times 10^{-7}$~\cite{Masjuan:2015cjl}, this cannot place stronger bounds.} This is complementary to first generation (electron) bounds from the EDMs of heavy atoms and molecules. Still, there would be possible ways to improve this study. They are beyond the scope of the present work, but we briefly comment on them in the following.
\\

Regarding the SMEFT operators, a possible extension would be an improved determination of nEDM bounds on $\mathcal{O}_{\ell equ,\ell edq}$ operators. There are different lines that could be pursued: considering non-vanishing $\mathcal{O}_{\ell equ}^{(3)}$ and $\mathcal{O}_{uW,uB,dW,dB}$ operators and employ the full RG equations~\cite{Pruna:2017tif}; computing the full two-loop calculation; improving the hadronic model (with a serious estimate of uncertainties). 
Also, one could estimate the impact on the same operators for the $\ell=\tau$ case. Here, the large-logs will become as important as hadronic effects, as they are $\propto m_{\ell}$, and the hadronic model might have to be improved up to higher scales. 
Very differently, it might be interesting to check the induced $\mathcal{O}_{\ell equ(dq)}^{11st}$ operators that might appear at two loops from $\mathcal{O}_{\ell equ(dq)}^{22st}$ and to check whether these might allow to improve the bounds derived here.
Finally, one might wonder about the $\mathcal{O}_{le}$ operator. As said, this does not produce an effect at LO in dilepton decays. In Dalitz decays, would be analogous (up to $i$ factor) to the $Z$-boson contribution, which we found negligible. For double Dalitz decays it might appear as a loop contribution, so we expect this small, with lepton EDMs presumably setting stronger bounds~\cite{Panico:2018hal} \\

Regarding additional decays, we did not discuss here the $\eta\to\mu^+\mu^-\pi^+\pi^-$ decay, especially in the \CPH{} scenario. Yet the latter has a larger BR than the leptonic one, the nEDM contribution would be very similar (for the \CPH{} scenario) to that in \cref{sec:nedmh} up to an  $\alpha^{-1}K^2$ factor,\footnote{The $\pi^+\pi^-$ state is essentially the low-energy manifestation of the vector isovector current, which would result in a similar diagram modulo photon propagator and form factors.} which would result in stronger bounds. For the \CPL{} scenario, on turn, we expect too small asymmetries as it happens for the leptonic case. Overall, we do not expect---in principle---any \CP{} violation in these decays.
Finally, we did not discuss polarizations in the $\eta\to\pi^0\mu^+\mu^-$ decay, that might be interesting to analyze~\cite{Gatto:2016rae}, but are beyond the scope of this study.
\\

{\noindent{\textit{Acknowledgements}}}\\
\indent The author acknowledges C.~Gatto for stimulating this work and comments concerning REDTOP experiment, J.~Novotn{\'y} for comments on two-loop renormalization, J.~M.~Alarc{\'o}n and P.~Masjuan for discussions regarding hadronic models, B.~Kubis for discussions regarding $\sigma$ terms, A.~Pineda and A.~Pomarol for discussions concerning the two-loops leading logs and K.~Kampf for discussions at early stages of this work. The author acknowledges the support received from the Ministerio de Ciencia, Innovación y Universidades under the grant SEV-2016-0588, the grant 754510 (EU, H2020-MSCA-COFUND-2016), and the grant FPA2017-86989-P, as well as by Secretaria d’Universitats i Recerca del Departament d’Economia i Coneixement de la Generalitat de Catalunya under the grant 2017~SGR~1069. This work was also supported by the Czech Science Foundation (grant no. GACR 18-17224S) and by the project UNCE/SCI/013 of Charles University.

\appendix

\section{The form factors parametrization}\label{sec:TFF}

Here we describe the parametrizations employed for the TFFs appearing in \cref{eq:ggvertex}. Regarding the standard---\CP{} conserving---one, $F_{\eta\gamma^*\gamma^*}(q_1^2,q_2^2)$, we employ the approach described in Refs.~\cite{Masjuan:2017tvw,Sanchez-Puertas:2017sih} and stick to the simplest parametrization that implements precisely the low-energy behavior and respects the high-energy one~\cite{Masjuan:2017tvw}\footnote{We checked that higher order approximants did not change significantly our results, so we stick to this for simplicity.}
   \begin{equation}
     F_{\eta\gamma^*\gamma^*}(q_1^2,q_2^2) = F_{\eta\gamma\gamma}\frac{\Lambda^2}{\Lambda^2 -q_1^2 -q_2^2},  
   \end{equation}
where, $F_{\eta\gamma\gamma} =  0.2738~\textrm{GeV}^{-1}$ and $\Lambda = 0.724~\textrm{GeV}$~\cite{Masjuan:2017tvw,Escribano:2015nra}, except when imaginary parts are relevant, which we postpone to the end of this section. For the \CP{}-violating form factors there is of course no theoretical knowledge as they are speculative and its microscopic origin is unknown. In the following, we assume the high-energy behavior from Ref.~\cite{Kroll:2016mbt}, implying that $F_{\eta\gamma^*\gamma^*}^{C\!P1}(-Q_1^2,-Q_2^2)$ and $-\bar{Q}^2F_{\eta\gamma^*\gamma^*}^{C\!P2}(-Q_1^2,-Q_2^2)$ behave as $\bar{Q}^{-2}$, where $2\bar{Q^2} = Q_1^2+Q_2^2$.
Thereby we choose
  \begin{equation}
    F_{\eta\gamma^*\gamma^*}^{C\!P1}(q_1^2,q_2^2) =  \frac{\epsilon_1\Lambda^2 F_{\eta\gamma\gamma}}{\Lambda^2 -q_1^2 -q_2^2}, \
    F_{\eta\gamma^*\gamma^*}^{C\!P2}(q_1^2,q_2^2) =   \frac{-2\epsilon_2\Lambda^2 F_{\eta\gamma\gamma}}{(\Lambda^2 -q_1^2 -q_2^2)(\Lambda_H^2 -q_1^2 -q_2^2)}.
  \end{equation}
We take the same value for $\Lambda$ as before and introduce $\Lambda_H=1.5$~GeV inspired by heavier resonances (results are rather stable upon varying these masses).

If only the imaginary parts are relevant for the asymmetry, we employ the TFF described in Ref.~\cite{Masjuan:2015cjl} instead. This reads $F_{P\gamma^*\gamma}(s) = F_{P\gamma\gamma}[c_{P\rho} G_{\rho}(s) + c_{P\omega} G_{\omega}(s) + c_{P\phi} G_{\phi}(s)]$, with $G_{\rho,\phi}(s)$ Breit-Wigner functions, and the intermediate $\pi\pi$ rescattering modeled as in Refs.~\cite{GomezDumm:2000fz,Dumm:2013zh} through 
  \begin{equation}
\label{eq:rho}
G_{\rho}(s) = \frac{M_{\rho}^2}{   M_{\rho}^2 - s +\frac{s M_{\rho}^2}{96\pi^2F_{\pi}^2}\left( \ln\left(\frac{m_{\pi}^2}{\mu^2}\right)  +\frac{8m_{\pi}^2}{s} -\frac{5}{3} - \sigma(s)^3 \ln\left(\frac{\sigma(s)-1}{\sigma(s)+1}\right)   \right)   }
  \end{equation}
and $c_{\eta\{\rho,\omega\phi\}} = \{9,1,-2  \}/8$, which is---effectively---similar to the description in Ref.~\cite{Hanhart:2013vba}.

\section{Polarized muon decay}\label{sec:polmudec}

In the effective Fermi theory, and using polarized spinor sums, we find for the $\mu^{\pm}\to e^{\pm}(k) \nu_{\mu}(q_1)\nu_{e}(q_2)$ decay amplitude
  \begin{equation}
    \big\vert\mathcal{M}\left(\mu^{\pm},\lambda \SLV{n}\right)\big\vert^2 = 64G_F^2 k_{\alpha}( p_{\beta} \pm \lambda  m_{\mu}n_{\beta} )q_1^{\alpha}q_2^{\beta}.
  \end{equation}
Including phase-space and integrating over the neutrino spectra (we employ the muon rest frame), the result above reads\footnote{In the second line, the result for integration over $d\Omega dx$ has been employed, that introduces $\epsilon=m_e^2\left[m_e^2(m_\mu^2-m_e^2)^2 +6m_{\mu}^6 +2m_e^2m_{\mu}^4\left(1 +6\ln(m_e/m_{\mu}) \right)\right](m_e^2 +m_{\mu}^2)^{-4}$. This is, modulo radiative correction effects, the SM result from Ref.~\cite{PDGmuon} and implemented in {\textsc{Geant4}}.}
   \begin{align}
      \frac{d\Gamma(\mu^{\pm},\lambda \SLV{n})}{dxd\Omega} &{}= 
      \frac{m_{\mu}}{8\pi^4}W_{e\mu}^4G_F^2\beta x^2 n(x,x_0)\left[  
                1 \mp \lambda b(x,x_0) \SLV{\beta}\cdot\SLV{n}  
      \right],\\\label{eq:mudecayexact}
      d\textrm{BR}(\mu^{\pm},\lambda \SLV{n}) &{}= 
      \frac{d\Omega}{4\pi} \frac{2x^2\beta}{1-2\epsilon}   n(x,x_0)\left[  
                1 \mp \lambda b(x,x_0) \SLV{\beta}\cdot\SLV{n}  
      \right]dx,
   \end{align}
with $n(x,x_0)=(3-2x-x_0^2/x)$ and $n(x,x_0)b(x,x_0)=2- 2x -\sqrt{1-x_0^2}$. Above, $W_{e\mu}=(m_{\mu}^2+m_e^2)/2m_{\mu}$ is the maximum positron energy, $x=E_e/W_{e\mu}$ the reduced positron energy, $x_0=m_e/W_{e\mu}$ the minimum reduced positron energy and $\beta=\sqrt{1-x_0^2/x^2}$ has the usual meaning. Typically, the approximation $m_e/m_{\mu}\to 0$ is employed, that results in the simpler expression
   \begin{equation}\label{eq:mudecayapp}
      d\textrm{BR}(\mu^{\pm},\lambda \SLV{n})= 
      \frac{d\Omega}{4\pi} n(x)  \left[  
                1 \mp \lambda b(x,x_0) \SLV{\beta}\cdot\SLV{n}  
      \right]dx,
\end{equation}
with $x=2E_e/m_{\mu}$, $n(x)=2x^2(3-2x)$ and $b(x)=(1-2x)/(3-2x)$.

\section{Results in Dalitz decays}\label{sec:ddres}

The resulting amplitudes for our \CP{}-violating scenarios read
   \begin{align}
     i\mathcal{M}^{\textrm{\CP}_{\textrm{H}}} =& -i e^3 q^{-2} F_{\eta\gamma\gamma^*}^{C\!P1}(q^2)\varepsilon^{\mu*} 
                \left(g_{\mu\nu}(k\cdot q) -q_{\mu}k_{\nu}\right) (\bar{u}\gamma^{\nu}v), \\
      i\mathcal{M}^{C\!P_{\textrm{HL}}} =& - ie \mathcal{C} \varepsilon_{\mu}^*\left[  
               \frac{\bar{u}(\gamma^{\mu}\slashed{k} +2p_-^{\mu})v}{2k \cdot p_-} 
               +\frac{\bar{u}(\gamma^{\mu}\slashed{k} -2p_+^{\mu})v}{2k \cdot p_+} 
               \right].
   \end{align}
Their interference with the SM amplitude in \cref{eq:dalitzsm} ($\operatorname{Int}_{X}= 2\operatorname{Re}\mathcal{M}^{\textrm{SM}}\mathcal{M}^{X}$) yield
  \begin{align}
      \operatorname{Int}_{C\!P_{\textrm{H}}} =& -\frac{1}{4} e^6 2\operatorname{Im} F_{\eta\gamma\gamma^*}(s) 
                        F_{\eta\gamma\gamma^*}^{C\!P1*}(s) (m_{\eta}^2-s)^2 s^{-1} \nonumber \\ & \quad
                        \times\left[  
                               \sqrt{1-\beta^2}\sin{\theta}(\lambda n_y +\bar{\lambda}\bar{n}_y) 
                               -\cos{\theta}(\lambda n_z+\bar{\lambda}\bar{n}_z)  
                        \right],\\
      \operatorname{Int}_{C\!P_{\textrm{LH}}} =& - \frac{1}{4} \frac{4e^4\mathcal{C}}{s(1-y^2)}
               \left[ \alpha_R\operatorname{Re}F_{\eta\gamma\gamma^*}(s) -\alpha_I\operatorname{Im}F_{\eta\gamma\gamma^*}(s)\right],
  \end{align}
where we introduced the following coefficients
   \begin{align}
     \alpha_R =&{} 
     \beta_{\mu}\sin{\theta}\left\{ n_x\left[ (m_{\eta}^2-s)
          \left[ \sqrt{s}\bar{n}_z(\beta_{\mu}-\cos{\theta}) +2m_{\mu}\sin{\theta}\bar{n}_y \right] +2\beta_{\mu}\bar{n}_zs^{3/2} 
     \right] \right.\nonumber\\ & \left.
   + \bar{n}_x\left[ (m_{\eta}^2-s)
          \left[ \sqrt{s}n_z(\beta_{\mu}+\cos{\theta}) -2m_{\mu}\sin{\theta}n_y \right] +2\beta_{\mu} n_zs^{3/2} \right] \right\}\lambda\bar{\lambda},\\
      \alpha_I =&{}
      2m_{\mu}(m_{\eta}^2 -s)\left[ -\lambda n_z(\beta_{\mu}\sin^2\theta +2\cos{\theta}) + \bar{\lambda}\bar{n}_z(\beta_{\mu}\sin^2\theta -2\cos{\theta}) \right] \nonumber\\ &
      -\lambda\sqrt{s}\sin{\theta} n_y\left[ (m_{\eta}^2 -s)(\beta_{\mu}\cos{\theta} -2) +\beta_{\mu}^2(m_{\eta}^2 -3s) \right] \nonumber\\ &
      +\bar{\lambda}\sqrt{s}\sin{\theta} \bar{n}_y\left[ (m_{\eta}^2 -s)(\beta_{\mu}\cos{\theta} +2) -\beta_{\mu}^2(m_{\eta}^2 -3s) \right].
   \end{align}
Finally, we give here the analytic results for the asymmetries in terms of the phase space integral
   \begin{align}
      A_{L\gamma}^H &{}= -\frac{1}{N}\int  \frac{\operatorname{Im}\tilde{F}(s)\tilde{P}^*_1(s)}{12s}(1-x_{\mu})^3\beta_{\mu}\sqrt{1-\beta_{\mu}^2} \ ds, \\
      A_{L\gamma}^{HL} &{}=  \frac{\tilde{\mathcal{C}}}{N}  \int \frac{\sqrt{x_{\mu}}( 1 -x_{\mu})}{3s\beta_{\mu}}
                       \operatorname{Im}F_{\eta\gamma\gamma^*}(s) 
                       \left[ 2(1-x_{\mu}) - \beta_{\mu}^2(1-3x_{\mu}) \right]\left( 1- \sqrt{1-\beta_{\mu}^2} \right) ds, \\
      A_{L}^{HL} &{}= -2\frac{\tilde{\mathcal{C}}}{N} \int \frac{(1 -x_{\mu})^2}{3\pi s \frac{m_{\eta}}{m_{\mu}}}
                       \operatorname{Im}F_{\eta\gamma\gamma^*}(s) 
                       \left[ 2 +(\beta_{\mu} -\beta_{\mu}^{-1})\ln\left( \frac{1+\beta}{1-\beta} \right) \right] ds, \\
      A_{TL}^{HL} &{}= \frac{\tilde{\mathcal{C}}}{N} \int \frac{\sqrt{x_{\mu}}( 1 -x_{\mu}^2)}{18s}
                       \operatorname{Re}F_{\eta\gamma\gamma^*}(s) \beta_{\mu} \left( 1- \sqrt{1-\beta_{\mu}^2} \right) ds, \\
      A_{T}^{HL} &{}= - \frac{\tilde{\mathcal{C}}}{N} \int \frac{(1 -x_{\mu})^2}{18s \frac{m_{\eta}}{m_{\mu}}}
                       \operatorname{Re}F_{\eta\gamma\gamma^*}(s) 
                       \left[ 2 +(\beta_{\mu} -\beta_{\mu}^{-1})\ln\left( \frac{1+\beta}{1-\beta} \right) \right] ds,  
   \end{align}
where we have introduced the common paremeter $\tilde{\mathcal{C}} =\mathcal{C}/(e^2 m_{\eta} F_{\eta\gamma\gamma})$\footnote{From \cref{eq:mathcalC}, $\tilde{\mathcal{C}}= (1.142(c_{\ell equ}^{(1)2211} +c_{\ell edq}^{2211}) -1.726 c_{\ell edq}^{2222})\times 10^{-4}$.} and
   \begin{equation}
      N = \frac{1}{3\pi} \int \frac{1}{s}|\tilde{F}_{\eta\gamma\gamma^*}(s)|^2 (1-x_{\mu})^3 \beta_{\mu} (3 -\beta_{\mu}^2) ds.
   \end{equation}
The latter is, up to an $\alpha$ factor, the $dy d_{e^{\pm}}$-integrated version of \cref{eq:dalitzpolsm}.

\section{$Z$ boson contribution to Dalitz decay}\label{sec:Zboson}

In the SM, parity-violating contributions arise from an intermediate $Z$-boson state,
   \begin{equation}
      i\mathcal{M}^Z = -i\frac{e G_F}{\sqrt{2}} \epsilon_{\mu\nu\rho\sigma}k^{\nu}q^{\sigma}\varepsilon^{\mu*} 
                      (\bar{u}\gamma^{\rho}[(1 -4\sin^2\theta_w) +\gamma^5]v)F_{\eta\gamma Z^*}(q^2),
   \end{equation}
where $\sqrt{2}G_F = g^2/(4m_W^2)$. The term without the $\gamma^5$ is analogous to the QED result modulo form factor details and the $s^{-1}$ factor; the $\gamma^5$ term induces a parity-violating interference with the QED contribution. Particularly,\footnote{Though at this step it cannot be compared to the results in Ref.~\cite{Bernabeu:1998hy}, that uses a different frame, we compared intermediate steps against their result in Eq.~(9). We found agreement up to a minus sign (we remark that they calculate the $\mu^+$ polarization, lacking the necessary terms to compare to our full polarization amplitude).} 
   \begin{multline}
      2\operatorname{Re} \mathcal{M}^{\textrm{SM}}\mathcal{M}^{Z,\slashed{P}} = 
           \frac{1}{4} \frac{e^4 G_F}{\sqrt{2}}(m_{\eta}^2 -s^2)^2 \beta_{\mu} \Big[ 
              \Big\{ (n_z+\bar{n}_z)(1+\cos^2\theta) - \\ (n_y+\bar{n}_y)\sin\theta\cos\theta\sqrt{1-\beta_{\mu}^2} \Big\} 
                   \operatorname{Re}F_{P\gamma\gamma^*}F_{P\gamma Z^*}^*  - 
              \sin\theta\Big\{ (n_x\bar{n}_y+n_y\bar{n}_x)\sin\theta \\ -(n_x\bar{n}_z+n_z\bar{n}_x)\sqrt{1-\beta_{\mu}^2}\cos\theta  \Big\} \sin\theta
                   \operatorname{Im}F_{P\gamma\gamma^*}F_{P\gamma Z^*}^*
           \Big].
   \end{multline} 
Again, the full decay width can be obtained to be
   \begin{align}
      \frac{d\Gamma}{\Gamma_{\gamma\gamma}} =& \frac{G_F}{8\sqrt{2}\pi^2}(1-x_{\mu})^3\beta_{\mu} ds dy de^{\pm} \Big[
         \alpha_R^Z \operatorname{Re}\tilde{F}_{P\gamma\gamma^*}\tilde{F}_{P\gamma Z^*}^*  +
         \alpha_I^Z \operatorname{Im}\tilde{F}_{P\gamma\gamma^*}\tilde{F}_{P\gamma Z^*}^* 
      \Big],\\
       &\alpha_R^Z = (\bar{b}\bar{\beta}_z - b\beta_z)(1+\cos^2\theta) - (\bar{b}\bar{\beta}_y -b\beta_y)\sin\theta\cos\theta\sqrt{1-\beta_{\mu}^2},\\
       &\alpha_I^Z = (\beta_x\bar{\beta}_y+\bar{\beta}_x\beta_y)\sin^2\theta - 
                    (\beta_x\bar{\beta}_z+\bar{\beta}_x\beta_z)\sin\theta\cos\theta\sqrt{1-\beta_{\mu}^2}.
   \end{align}
With these results at hand, one finds that the $Z$-boson parity-violating contribution results in a non-vanishing asymmetry
   \begin{equation}
      A_{L} = -\bar{A}_{L} = \frac{1}{N\alpha} \frac{G_F}{18\sqrt{2}\pi^2}\int ds (1-x_{\mu})^3\beta_{\mu}^2
                                       \operatorname{Re}\tilde{F}_{P\gamma\gamma^*}\tilde{F}_{P\gamma Z^*}^*. 
   \end{equation}
Regarding the form factor, we take its normalization from $\chi$PT at NLO (see Refs.~\cite{Feldmann:1999uf,Escribano:2015yup})\footnote{With this, we obtain $F_{\eta\gamma Z^*}(0,0) = 0.074(6)~\textrm{GeV}^{-1}$ and $F_{\eta'\gamma Z^*}(0,0) = 0.19(1)~\textrm{GeV}^{-1}$.}
   \begin{equation}
      F_{\eta\gamma Z} = \frac{1}{4\pi^2\sqrt{3}}
      \frac{F_{\eta'}^0c^{8}_{Z} +2\sqrt{2}F_{\eta'}^8c^{0}_{Z}}{F_{\eta}^8F_{\eta'}^0 -F_{\eta}^0F_{\eta'}^8},
   \end{equation}
where $c^{8}_{Z} = \frac{1}{2}(1 \! + \!K_2^{8T_3}\! -4\sin^2\!\theta_w c^{8}_{\gamma} )$, $c^{0}_{Z} = (1 \! + \! K_2^{0T_3} \! + \!K_1 -2\sin^2 \!\theta_w c^{0}_{\gamma} )$, $ K_2^{8T_3} = K_2 (5m_{\pi}^2 -4m_K^2)$, and $K_2^{0T_3} = K_2 (m_K^2 +m_{\pi}^2)/2$ and the mixing parameters are taken from \cite{Escribano:2015yup}.
Finally, for its $q^2$-dependence, we take a TFF analogous to that in \cref{sec:TFF} with $\Lambda=0.57(7)$~GeV and $0.90(4)$~GeV for the $\eta$ and $\eta'$.  This results from averaging the values that would be obtained from a BL-interpolation formula~\cite{Brodsky:1981rp} and from a resonance saturation approach with weights given by mixing parameters similar to Refs.~\cite{Landsberg:1986fd,Feldmann:1999uf,Escribano:2015nra}. As a result, we find $A_{L} =6(1)\times 10^{-7}$,\footnote{This might be compared to Ref.~\cite{Bernabeu:1998hy} results. Checking intermediate steps, we confirmed their results except for an overall sign. Still, we find 2 orders of magnitude supression. This is due to the relevant scale in the problem ($2m_{\mu}$ rather than $m_{\eta}$) and the resolution power function $b(x)< 1$.} which is irrelevant for the expected statistics at REDTOP, and would require of the order of $10^{16}$ $\eta$ mesons for its observation.

\section{The nucleon scalar form factors}\label{app:scalarff}

\begin{figure}\centering
   \includegraphics[width=0.5\textwidth]{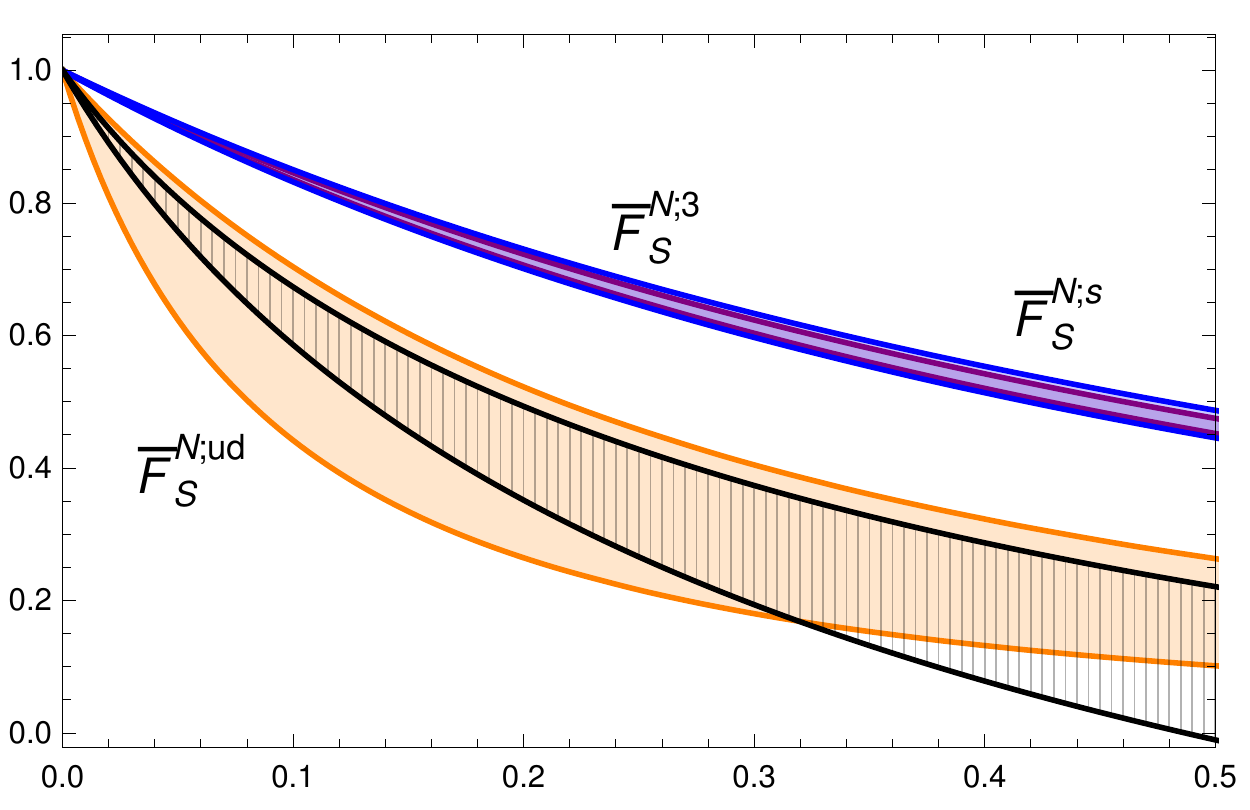}
   \caption{Comparison of the half-width estimate for the---normalized---isovector form factor (orange band) to that in Ref.~\cite{Alarcon:2017ivh}. We also include the isovector (blue) and strange (purple). \label{fig:sff}}
\end{figure}

In this section, we introduce the nucleon scalar form factors $F_S^{N;q}(q^2) \equiv \bra{N_{p'}} \bar{q}q \ket{N_p}$. At $q^2=0$, these are related to the $\sigma$-terms. In the following, we average theoretical (if available) and lattice results from Ref.~\cite{Varnhorst:2015hfk} (enlarging errors if necessary). Regarding the theoretical input, for the isoscalar, $\sigma_{ud}^N \equiv \bra{N_{p}} \hat{m}(\bar{u}u + \bar{d}d) \ket{N_p}$, we take those obtained\footnote{The relation to this matrix element is outlined in Ref.~\cite{Crivellin:2013ipa}.} in Ref.~\cite{Hoferichter:2015dsa} and the $\sigma_{\pi N}$ result in~\cite{Alarcon:2011zs}; for the isovector, $\sigma_{3}^N \equiv \bra{N_{p}} \hat{m}(\bar{u}u - \bar{d}d) \ket{N_p}$, we use both, those that can be obtained from Ref.~\cite{Hoferichter:2015dsa},  and those appearing in Ref.~\cite{Fernando:2018jrz}\footnote{In doing so, we use the value arising from a $\chi$PT-based analysis $m_d/m_u=0.553(43)$~\cite{Tanabashi:2018oca} rather than lattice.}; for the strange one, $\sigma_{3}^s = \bra{N_{p}} m_s\bar{s}s  \ket{N_p}$, we restrict to the lattice results due to the large theoretical uncertainties. This results in
  \begin{align}
    \sigma_{ud}^{p,n} = 53(16), \ 
    \sigma_{3}^{p} = 2.9(1.0),  \
    \sigma_{3}^{n} = -2.9(0.6),  \
    \sigma_{s}^{p,n} = 54(5),
  \end{align}
in MeV units. Regarding the $q^2$-dependency, we use the half-width rule~\cite{Arriola:2012vk}, which proved to provide excellent estimates for the form factors. Since at high energies $F_S^{N;q}(-Q^2)\sim Q^{-6}$~\cite{Bali:2018qus,Alabiso:1974ye}, we use three resonances. Following \cite{PDGscalar}, we employ for the isoscalar channel $f_0(500)$, $f_0(1370)$, $f_0(1500)$, for the isovector $a_0(980)$, $a_0(1450)$, $a_0(1950)$, and for the strange one $f_0(980)$, $f_0(1500)$, $f_0(1710)$. In the following, we provide the central value for the required form factors
  \begin{align}
     F_S^{p;u(d)}& = \frac{7.57}{(1 +\frac{q^2}{m_{f_1}^2})(1 +\frac{q^2}{m_{f_3}^2})(1 +\frac{q^2}{m_{f_4}^2})} \pm  \frac{0.41}{(1 +\frac{q^2}{m_{f_2}^2})(1 +\frac{q^2}{m_{f_4}^2})(1 +\frac{q^2}{m_{f_5}^2})},\\
     F_S^{n;u(d)}& =  \frac{7.57}{(1 +\frac{q^2}{m_{f_1}^2})(1 +\frac{q^2}{m_{f_3}^2})(1 +\frac{q^2}{m_{f_4}^2})} \mp \frac{0.41}{(1 +\frac{q^2}{m_{f_2}^2})(1 +\frac{q^2}{m_{f_4}^2})(1 +\frac{q^2}{m_{f_5}^2})},\\
     F_S^{N;s}& =\frac{0.57}{(1 +\frac{q^2}{m_{a_1}^2})(1 +\frac{q^2}{m_{a_2}^2})(1 +\frac{q^2}{m_{a_3}^2})},
  \end{align}
where $f_{1,2,3,4,5}=f_0(500,980,1370,1500,1710)$ and $a_i=a_0(980,450,1950).$ For the sake of illustration, we compare in \cref{fig:sff} the prediction for the (normalized) isoscalar form factor to that in Ref.~\cite{Alarcon:2017ivh}, obtaining an excellent agreement (note that Ref.~\cite{Alarcon:2017ivh} provides a reasonable estimate up to energies around $0.5$~GeV).

\bibliographystyle{apsrev4-1}
\bibliography{references}

\end{document}